\documentclass[doublecol]{epl2}
\usepackage[T1]{fontenc}
\usepackage[latin9]{inputenc}
\usepackage{amsbsy}
\usepackage{graphicx}
\usepackage{amssymb}
\usepackage{amsmath}

\DeclareMathOperator{\dist}{dist}
\DeclareMathOperator{\Real}{Re}

\title{Do Small Worlds Synchronize Fastest?}

\author{Carsten Grabow\inst{1} \and Steven Hill\inst{2} \and Stefan Grosskinsky\inst{2} \and Marc Timme\inst{1}}

\shortauthor{C. Grabow, S. Hill, S. Grosskinsky and M.Timme}

\institute{
\inst{1} Network Dynamics Group, Max Planck Institute for Dynamics and Self-Organization,
37073 G\"ottingen, Germany\\
\inst{2}Centre for Complexity Science, University of Warwick, Coventry CV4
7AL, UK
}


\abstract{
Small world networks interpolate between fully regular and fully random
topologies and simultaneously exhibit large local clustering as well
as short average path length. Small world topology has therefore been
suggested to support network synchronization. Here we study the asymptotic
speed of synchronization of coupled oscillators in dependence on the
degree of randomness of their interaction topology in generalized Watts-Strogatz ensembles. We find that networks
with fixed in-degree synchronize faster the more random they are,
with small worlds just appearing as an intermediate case. For any generic network ensemble, if synchronization speed is at all extremal 
at intermediate randomness, it is slowest in the small
world regime. This phenomenon occurs for various types
of oscillators, intrinsic dynamics and coupling schemes. 
}

\begin{document}

\maketitle

Synchronization dominates the collective dynamics of many physical
and biological systems \cite{SynchBook, Arenas:2008p1192}. It might be both advantageous
and desired, for instance in secure communication \cite{Kanter:2002p2846},
or detrimental and undesired, as during tremor in patients with Parkinson
disease or during epileptic seizures \cite{Maistrenko:2004p1453,Netoff:2004p241}.
Therefore, a broad area of research has emerged \cite{Strogatz01, Nishikawa:2003p656,Pecora:1998p265}, 
determining under which conditions on the interaction strengths and topologies coupled units actually synchronize
and when they do not. In a seminal work essentially founding the science of complex network
theory, Watts and Strogatz \cite{WattsStrogatz98} suggested that
a small world topology of a network is particularly supportive of
synchronization because small worlds exhibit high local clustering and simultaneously low average path length.
Indeed, several detailed studies support this view by showing that 
at fixed coupling strength
small world networks tend to already synchronize at lower connectivity than many other classes of networks \cite{WattsStrogatz98, BarahonaPecora}; small worlds also more easily exhibit self-sustained activity \cite{Roxin:2004p2947}.

These results suggest some key properties about the topological influence on the network synchronizability, i.e. the capability of a network to synchronize at all, but do not tell much about the speed of synchronization given that a network synchronizes in principle.

For any real system, however, it equally matters how fast the units synchronize
or whether the network interactions fail to coordinate the units'
dynamics on time scales relevant to the system's function (or dysfunction), cf. \cite{Zumdieck:1990p2850,Zillmer:2007p2849,Jahnke:2008p2847,Zillmer:2009p2851}.
Yet this question is far from being understood and currently under
active investigation \cite{Timme04,Timme06,Timme:2006p292,Qi:2008p1201,Qi:2008p1198}.
In particular it is largely unknown how fast small worlds synchronize,
an astounding fact given the seminal work on small world networks
\cite{WattsStrogatz98} published more than a decade ago.

In this Letter we study the speed of synchronization in generalized Watts-Strogatz ensembles and systematically compare the small world regime to more regular and more random topologies.
We find that small worlds synchronize faster than regular networks
but still orders of magnitude slower than fully random networks. The observed
increase of synchronization
speed with randomness might be attributed \cite{WattsStrogatz98,Timme06} to the simultaneous decrease of the
average path length between two units in the network. We therefore
compare ensembles of networks where the degree of randomness varies
from completely regular to completely random such that the average path
length stays constant.
Here we find that networks synchronize slowest in the small world regime. Within the entire model class, these results hold for any generic ensemble, i.e. synchronization speed may be intermediate or slowest but is never fastest in the small world regime.
This phenomenon occurs across many distinct systems, including phase oscillators,
higher-dimensional periodic and chaotic systems coupled diffusively
as well as neural circuits with inhibitory delayed pulse-coupling.

Consider $N$ Kuramoto oscillators \cite{Acebron:2005p293} that interact
on a directed network. The dynamics of phases $\theta_{i}(t)\in\mathbb{S}^{1}=2\pi \mathbb{R}/\mathbb{N}$ of oscillators $i$ with time $t$ satisfy \begin{equation}
\frac{d\theta_{i}}{dt}=\omega+\sum_{j}J_{ij}\sin(\theta_{j}-\theta_{i})\hspace{0.5cm}\,\mbox{for}\, i\in\{1,...,N\}\ ,\label{eq:Kuramoto}\end{equation}

\noindent where $\omega$ is the natural frequency of the oscillators, $J_{ij}=J/k$ is the coupling strength between two units and $k$ is the number of in-links to a unit. 
To analyze the purely
topological impact on the synchronization times, we study the network
dynamics in its simplest setting: we consider strongly connected networks with fixed in-degree $k$ and homogeneous total input coupling
strengths such that full synchrony is achieved from sufficiently close initial conditions for all coupling strengths $J>0$ \cite{Timme:2006p292}.

As the synchronous periodic orbit analyzed is isolated in state space, the relaxation time continuously changes with possible inhomogeneities, so the qualitative results obtained below are generic and also hold in the presence of small heterogeneities, cf.~\cite{Denker:2004p2853}.

\begin{figure}[thbp]
\begin{centering}
\includegraphics[trim = .5mm .2mm .2mm .4mm, clip, width=80mm]{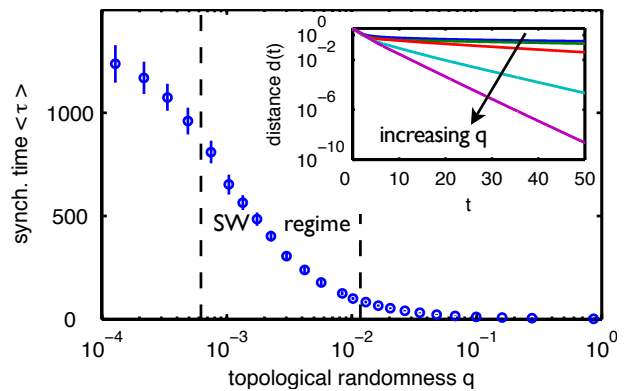}
\par\end{centering}

\caption{{\small (color online) Synchronization times (\ref{eq:distancedecay})
monotonically decrease with randomness in network ensembles with fixed
in-degree $k=20$ ($N=1000$ nodes, $J=1$, averages $\left\langle \tau\right\rangle $ over 100 realizations
of networks with random initial condition each; synchronization times ranging from $\left\langle \tau\right\rangle = 1.3$ ($q\rightarrow1$) to $\left\langle \tau\right\rangle = 1316$ ($q\rightarrow0$); error bars give standard
deviation). The small world regime {[}Eq. (\ref{eq:LCdefinitions}),
dashed vertical lines indicate bounds] appears not to be special at
all. Inset: Distance $d$ to the synchronous state (\ref{eq:distancedefinition})
decays exponentially with time $t$ after short transients for entire
range of randomness $q\in[0,1]$; lines provide single realizations
for $q\in\{0,0.008,0.04,0.2,1\}$. \label{fig:SWjustintermediate} }}

\end{figure}

To systematically investigate the sychronization process in dependence of
the topological randomness we first performed extensive numerical
simulations of the collective dynamics. We start with regular ring networks
where each unit receives directed links from its $k/2$ nearest neighbors on both sides.
Adapting the standard small
world model of Watts and Strogatz \cite{WattsStrogatz98} to directed
networks \cite{clustdef} we randomly cut the tail of each edge with
probability $q$ and rewire it to a randomly selected node (avoiding
double edges and self-loops). The small world regime is characterized
by a large clustering coefficient\footnote{{$C(q,k)$ denotes the actual divided by the possible number of
directed triangles containing a given node $i$, averaged over all $i$.}} $\left\langle C(q,k)\right\rangle $ and a small average path length\footnote{$L(q,k)$ denotes the length of the shortest directed path between a given
pair of nodes $(i,j)$, averaged over all $(i,j)$.}. $\left\langle L(q,k)\right\rangle$
Here $\left\langle .\right\rangle $ denotes averaging over network
realizations at given $q$ and $k$. To quantitatively fix the small
world regime we take
\begin{equation}
\frac{\left\langle L(q,k)\right\rangle }{L(0,k)}<0.5\quad\mbox{and}\quad\frac{\left\langle C(q,k)\right\rangle }{C(0,k)}>0.85\label{eq:LCdefinitions}
\end{equation}
throughout this study. The results below are not sensitive to a change
of these values. Starting each simulation from a random initial phase
vector drawn from the uniform distribution on $[0,\pi)^{N}$ shows
that synchronization becomes an exponential process after some short
transients (Fig.~\ref{fig:SWjustintermediate}, inset), for all fractions
$q\in (0,1]$ of randomness. Thus the distance \begin{equation}
d(t)=\max_{i,j}\dist(\theta_{i}(t),\theta_{j}(t))\label{eq:distancedefinition}\end{equation}
from the synchronous state decays as
\begin{equation}
d(t)\sim\exp(-t/\tau)\label{eq:distancedecay}\end{equation}
in the long time limit, where $\dist(\theta,\theta')$
is the circular distance between the two phases $\theta$ and $\theta'$
on $\mathbb{S}^{1}$.

The asymptotic synchronization time $\tau$ systematically depends
on the network topology (Fig.~\ref{fig:SWjustintermediate}): Regular ring networks ($q\rightarrow0$) are
typically relatively slow to synchronize. 

We find that increasing randomness $q$ towards the small world regime
induces shorter and shorter network synchronization times, with small
worlds synchronizing a few times faster than regular rings. 
Further increasing the randomness $q$ induces
even much faster synchronization, with fully random networks ($q\rightarrow1$)
synchronizing fastest (two orders of magnitude faster than small worlds in our examples).
Thus in network ensembles with fixed in-degree small worlds just occur intermediately during a monotonic increase of synchronization speed, but are not at all topologically optimal regarding their synchronization time.

This might be expected intuitively, also from studies about synchronizability \cite{WattsStrogatz98,BarahonaPecora}, and one is tempted to ascribe faster synchronization to a shorter average path length that results from increasing randomness.

We therefore first systematically studied the synchronization time for generalized Watts-Strogatz ensembles
of networks, specified by a function $k(q)$, where the average path length $\left\langle L\right\rangle $ is fixed while the
degree of randomness $q$ varies.\footnote{We choose an appropriate in-degree $k(q)$ for each given
randomness $q$ from numerically determined calibration curves such that
$\left\langle L(q,k(q))\right\rangle $ is fixed.}

\begin{figure}[htbp]
\begin{centering}
\includegraphics[trim = .5mm .2mm .2mm .2mm, clip, width=80mm]{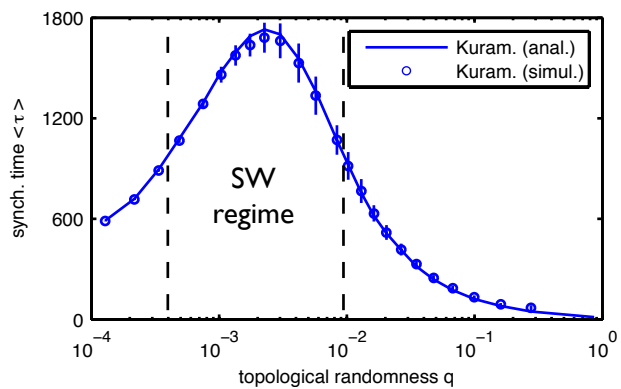}
\par\end{centering}

\caption{{\small (color online) Small worlds exhibit }\emph{\small slowest}{\small{}
synchronization in network ensembles with fixed average path length.
Here $N=1000$ and $\left\langle L\right\rangle =4$;
small-world region (\ref{eq:LCdefinitions}) is located between dashed
vertical lines.\label{fig:synchtimeLfixedKuramoto}}}

\end{figure}

We were surprised to find a non-monotonic behavior of synchronization
time with randomness (Fig.~\ref{fig:synchtimeLfixedKuramoto}). In
particular, networks with intermediate randomness in the small world
regime synchronize slowest. Analytical calculations
support this view. In the dynamics linearizing (\ref{eq:Kuramoto}) close
to the synchronous state (where $\theta_{i}(t)\equiv\theta_{j}(t)=:\theta(t)$)
phase perturbations $\varphi_{i}(t):=\theta_{i}(t)-\theta(t)$
evolve according to
\begin{equation}
\frac{d\varphi_{i}}{dt}=\sum_{j}\Lambda_{ij}\varphi_{j}\hspace{0.5cm}\,\mbox{for}\, i\in\{1,...,N\}.\label{linearmodel}
\end{equation}
Here the stability matrix coincides with the graph Laplacian defined as \begin{equation}
\Lambda_{ij}=J_{ij}(1-\delta_{ij})-J \delta_{ij} \label{laplacian}\end{equation}
and $\delta_{ij}$ is the Kronecker-delta. Close to every invariant trajectory the eigenvalue $\lambda_{2}$
of the stability matrix $\Lambda$ that is second largest in real
part dominates the asymptotic decay; therefore,  $\lambda_{2}$ here determines the asymptotic synchronization time via $\tau=-\frac{1}{\Real \lambda_{2}} $.
This feature was recently shown to hold more generally for network systems where the stability matrix is not necessarily proportional to the graph Laplacian \cite{Timme04, Almendral:2007p1211, Arenas:2008p1192}.

Determining the eigenvalues of the stability matrices of networks with fixed
average path lengths yields synchronization time estimates that well
agree with those found from direct numerical simulations, cf.~Fig.~\ref{fig:synchtimeLfixedKuramoto}.
This independently confirms that synchronization is indeed slowest
for small world networks.

\begin{figure}[htbp]
\begin{centering}
\includegraphics[width=80mm]{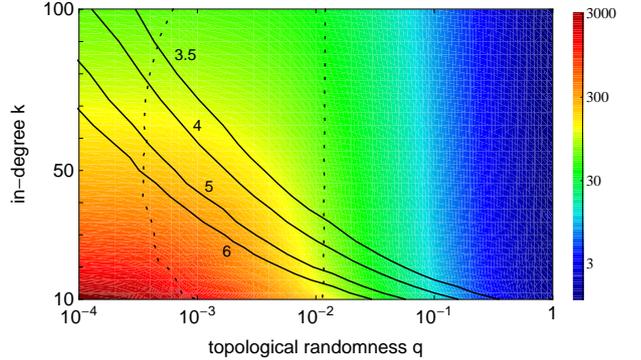} 
\par\end{centering}

\caption{\small (color online)
Nonlinear dependence of synchronization time on in-degree $k$ and
topological randomness $q$ indicates that no generic ensemble $k(q)$ exhibits
fastest synchronization in the small world regime (logarithmic color scale ranging from $\langle\tau\rangle =4606$
(dark red) to $\langle\tau\rangle =1.1$ (dark blue)). 
Solid lines indicate ensembles
of networks with fixed average path length from $\langle L\rangle=3.5$
(top) to $\langle L\rangle=6$ (bottom). The
small world regime (\ref{eq:LCdefinitions}) is located between the
dashed lines.\label{fig:synctimeqk} }

\end{figure}

How does synchronization speed vary with randomness for more general ensembles $k(q)$?
A systematic study of the synchronization time as a function of both
in-degree $k$ and randomness $q$ (Fig. \ref{fig:synctimeqk}) reveals
an interesting nonlinear dependence. Firstly, it confirms that for all
networks with fixed in-degree $k$ the synchronization time is monotonic
in the randomness $q$ and the small world regime at intermediate
randomness is not specifically distinguished. Secondly, the two-dimensional
function $\left\langle \tau(q,k) \right\rangle$ implies that ensembles of networks with fixed
path lengths all exhibit a non-monotonic behavior of the synchronization
time, with \emph{slowest} synchronization for intermediate randomness.

Thirdly, considering graph ensembles characterized by any other smooth function $k(q)$, $q\in[0,1]$, shows that this is a general phenomenon and the specific choice of an ensemble $k(q)$ is not essential.

In fact, as illustrated in Fig.~\ref{fig:synctimeqk}, for any generic network ensemble $k(q)$ (including ensembles with fixed in-degree and fixed path length as special choice)
the synchronization speed $\left\langle \tau(q,k(q))\right\rangle $ is either intermediate or slowest, but never
fastest at intermediate randomness, in particular in the small world regime.

\begin{figure}[htb]
\begin{centering}
\includegraphics[trim = .5mm .1mm 0mm .2mm, clip, width=80mm]{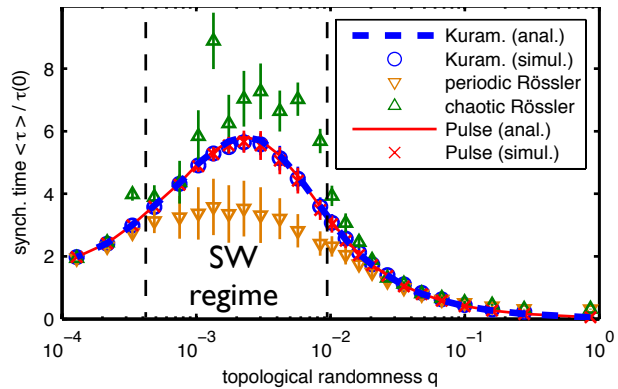}
\par\end{centering}

\caption{{\small (color online) Synchronization is slowest in the small world
regime for various oscillator types and coupling schemes (network parameters as in
Fig.~\ref{fig:synchtimeLfixedKuramoto}). Normalized
average synchronization times $\left\langle \tau\right\rangle /\tau(0)$
vs.~$q$ for Kuramoto, pulse-coupled and
periodic as well as chaotic R\"ossler oscillators (network topologies
as in Fig.~\ref{fig:synchtimeLfixedKuramoto}). Neurons with
delayed pulse-coupling: $I=1.01$, $\gamma=1$, $J=-0.2$, $\tau=0.1$; Diffusively coupled R\"ossler oscillators: $a=0.2$, $c=5.7$, periodic: $b=1.7$, $J=2$,
chaotic: $b=0.2$, $J=6$. The synchronization
times are determined by measuring the distances  $d=\max_{i,j}\{\big((x_{i}-x_{j})^{2}+(y_{i}-y_{j})^{2}+(z_{i}-z_{j})^{2}\big)^{1/2}\}$ (R\"ossler), and $d=\max_i |\delta_i |$ (pulse-coupled) and fitting (\ref{eq:distancedecay}).
\label{fig:ALLsynchtimes}}}
\end{figure}

Is this phenomenon restricted to the specific class of Kuramoto oscillators?
To answer this question, we explored the synchronization dynamics
of various kinds of oscillators coupled in different ways, and consistently
found qualitatively the same results. Specifically, in networks with fixed average
path length, synchrony is consistently fast for regular rings, fastest for completely
random networks, and slowest in the intermediate small world regime (Fig. \ref{fig:ALLsynchtimes}).

For instance, we tested networks of diffusively coupled three-dimensional R\"ossler
oscillators \cite{SynchBook} satisfying\begin{equation}
\begin{array}{l}
\dot{x}_{i}=-y_{i}-z_{i}+\sum_{j}J_{ij}(x_{j}-x_{i}),\\
\dot{y}_{i}=x_{i}+ay_{i},\\
\dot{z}_{i}=b+z_{i}(x_{i}-c),\end{array}\label{roessler}\end{equation}
for $i\in\{1,\ldots,N\}$ where $J_{ij}=J/k$ define the diffusive
coupling and the parameters $a$, $b$ and $c$ determine whether
the oscillators are intrinsically periodic or intrinsically chaotic. The above phenomenon persists for both periodic and chaotic
oscillators (Fig. \ref{fig:ALLsynchtimes}, triangles).

Moreover, we investigated the collective dynamics of pulse-coupled neural oscillators \cite{MIROLLO:1990p366,Jahnke:2008p2847} with membrane
potentials $V_{i}(t)$ satisfying \begin{equation}
\frac{dV_{i}}{dt}=I- \gamma V_{i}+\sum_{j=1;\, j\neq i}^{N}\sum_{m\in\mathbb{Z}}J_{ij}\delta\left(t-\left(t_{j,m}+\Delta\right)\right).\end{equation}
Here, each potential $V_{j}$ relaxes towards $I>1$ and is
reset to zero whenever it reaches a threshold at unity,
\begin{equation}
V_{j}(t^{-})=1\,\Rightarrow\, V_j (t):=0,\, t_{j,m}:=t,\,\mbox{and}\, m\mapsto m+1.\end{equation} 
At these times $t_{j,m}$ the neuron sends a pulse that after a delay
$\Delta>0$ changes the potential of post-synaptic neurons $i$ in an
inhibitory (negative) manner. This neural system allows analytic computation \cite{Timme06}
of an iterative map \begin{equation}
\delta_i (nT)=\sum_{j=1}^N A_{ij} \delta_j \big( (n{-}1)\, T\big)\ ,\quad n\in\mathbb{N},
\end{equation} 

for the perturbations $\delta_i (nT)$ of spike times
close to the synchronous orbit of period $T=(1/\gamma) \ln(1/(1-\gamma/I))$. For homogeneous coupling
($J_{ij}=-J/k$ for each existing connection) the elements of the
stability matrix $A$ are given by $A_{ij}=A_+=(\gamma J)/(k (I e^{- \gamma \Delta} +\gamma J))$
if there is a connection from $j$ to $i\neq j$, $A_{ii}=1-kA_{+}$ for the
diagonal elements and $A_{ij}=0$ otherwise, cf.~ \cite{Timme06}. 
As for the Kuramoto system, the prediction of synchronization times based on the eigenvalues
of the matrix $A$ well agrees with those obtained from direct numerical simulation
(Fig.~\ref{fig:ALLsynchtimes}, crosses and solid line).

These results confirm that, largely insensitive to the type of oscillators
(phase, multi-dimensional, neural), their intrinsic dynamics (periodic,
chaotic) and their coupling schemes (phase-difference, diffusive,
pulse-like), networks with fixed average path length consistently synchronize slowest
in the small world regime at intermediate randomness. Further numerical analysis
(not shown) indicates that also the entire nonlinear dependence (Fig.~\ref{fig:synctimeqk}) of the
synchronization time on $k$ and $q$ stays qualitatively the same for all these
different systems.

Hence, in general small worlds do not synchronize fastest. 
This holds for various oscillator types, intrinsic dynamics and coupling schemes: phase oscillators coupled via phase differences,
higher-dimensional periodic and chaotic systems coupled diffusively
as well as neural circuits with inhibitory delayed pulse-coupling.
In particular, small world topologies are not at all special and 
may synchronize orders of magnitude slower than completely
random networks.  So generically the small world regime can either exhibit slowest
synchronization or just exhibit no extremal properties regarding
synchronization times.

This phenomenon is rather unexpected given previous results on synchronization
and small world topology. 
For instance, the original work by
Watts and Strogatz, as well as later studies
\cite{WattsStrogatz98,BarahonaPecora,Nishikawa:2003p656}, 
indicate that small world topologies support network synchronization, in particular they synchronize at weaker coupling strength than analogous, appropriately normalized globally coupled systems.

Apart from small world properties, other topological features such as betweenness centrality, degree
heterogeneity or hierarchical organization have been suggested to control
whether or not a network actually synchronizes \cite{Motter:2005p76}.
Our results now highlight, that  
the speed of synchronization may vary several orders of magnitude, even if only the disorder in the topology changes. Synchronization speed thus serves as a 
key dynamic characteristic of oscillator networks, because even if a system synchronizes in principle, it might not in practice as the time
scales involved may be much longer than those relevant to the system's
function. For practical problems in real-world networks, such as preventing
synchrony in neural circuits \cite{Maistrenko:2004p1453}, or supporting
synchrony in communication systems \cite{Kanter:2002p2846}, it is thus
essential to further systematically investigate how additional features, such 
as heterogeneities \cite{Kurths2006} or non-standard degree distributions \cite{Qi:2008p1198}, impact synchronization speed.

\begin{acknowledgments}
Supported by the BMBF under grant no. 01GQ0430, by a grant of the
Max Planck Society to MT and by EPSRC grant no. EP/E501311/1 to
SH and SG. 
\end{acknowledgments}

\end{document}